\documentclass[twocolumn,showpacs,preprintnumbers]{revtex4}

\input{epsf}

\def\gev{\mbox{GeV}}

\def\tev{\mbox{TeV}}

\def\CQG{{\it Class. Quantum Gravity} }

\def\GRG{{\it Gen. Relativity and Gravitation} }

\def\LNC{{\it Lett. Nuovo Cimento} }

\def\MPL{{\it Mod. Phys. Lett.} }

\def\NAT{{\it Nature} }

\def\NP{{\it Nucl. Phys.} }
\def\PL{{\it Phys. Lett.} }
\def\PMAG{{\it Philos. Mag.} }
\def\PR{{\it Phys. Rev.} }
\def\PRL{{\it Phys. Rev. Lett.} }
\def\PRTS{{\it Physics Reports} }

\def\frac#1#2{{\textstyle{{#1}\over {#2}}}}

\def\lsim{\mathrel{\rlap{\lower4pt\hbox{\hskip1pt$\sim$}}
    \raise1pt\hbox{$<$}}}
\def\gsim{\mathrel{\rlap{\lower4pt\hbox{\hskip1pt$\sim$}}
    \raise1pt\hbox{$>$}}}
\def\sqr#1#2{{\vcenter{\vbox{\hrule height.#2pt
         \hbox{\vrule width.#2pt height#1pt \kern#1pt
         \vrule width.#2pt}
         \hrule height.#2pt}}}}

\def\beq{\begin{equation}}
\def\eeq{\end{equation}}
\def\beqa{\begin{eqnarray}} 
\def\eeqa{\end{eqnarray}}

\begin{document}

\preprint{DF/IST-5.2002}

\title{Ultracold neutrons, quantum effects of gravity and the Weak 
Equivalence Principle}

\author{O. Bertolami$^1$ and F.M. Nunes$^2$}

\affiliation{$^1$ Departamento de F\'{\i}sica, Instituto Superior T\'ecnico, 
Av. Rovisco Pais 1, 1049-001 Lisboa, Portugal}

\affiliation{$^2$ Universidade Fernando Pessoa, Pra\c{c}a 9 de Abril, 
4200 Porto, Portugal}

\affiliation{E-mail addresses: filomena@wotan.ist.utl.pt; 
orfeu@cosmos.ist.utl.pt}

\vskip 0.5cm

\date{\today}

\begin{abstract}
We consider an extension of the recent experiment with ultracold neutrons 
and the quantization of its vertical motion in order to test the Weak 
Equivalence  Principle. We show that an improvement on the energy resolution 
of the experiment may allow to establish a modest limit to the 
Weak Equivalence 
Principle and on the gravitational screening constant.
We also discuss the influence of a possible new interaction of Nature.

\vskip 0.5cm

\end{abstract}


\maketitle 
Quite recently, Nesvizhevsky and collaborators have reported on the 
evidence of quantization of the vertical motion of neutrons bound by the 
gravitational field of the Earth \cite{Nesvizhevsky}. Even though 
the conceptual 
aspects of the experiment were discussed back in the seventies 
\cite{Luschikov}, 
concrete steps towards the goal of the final experiment were realized 
more recently \cite{Nesvizhevsky1}. 
The experiment consists in allowing ultracold neutrons generated by a 
source at the Institute Laue-Langevin reactor (Grenoble) to fall towards 
a horizontal mirror due to the influence of the Earth's gravitational field. 
This potential confines the motion of the neutrons 
which no longer move continuously in the vertical 
direction, but rather jump from one height to another, as predicted by quantum 
mechanics. This impressive experiment complements the Collela, Overhauser 
and Werner experiment \cite{Collela} where neutrons were split and let 
interfere with the gravitational field, even though in that situation 
the neutrons were not quantum states. 
In the Grenoble experiment \cite{Nesvizhevsky}, the minimum energy 
of the detected neutrons is $1.4 \times 10^{-12}~eV$, corresponding to a 
vertical velocity of $1.7~cm/s$. In that work it is stated that a more intense 
beam and an enclosure mirrored on all sides could lead to an improved energy 
resolution, down to $10^{-18}~eV$, if the neutron can be kept confined 
all its lifetime. In  Ref. \cite{Nesvizhevsky} besides the ground state, 
three excited states were determined, although with reduced accuracy.

An earlier theoretical work \cite{Bertolami1} suggested the use of a version 
of the Collela, Overhauser and Werner experiment for establishing bounds 
on the parameters of a possible new interaction of Nature.
In this work however, 
we shall show that an upgraded version of the experiment of 
Nesvizhevsky and collaborators does not yield any
bound on the strength of new interaction of Nature and 
explore various implications
of a highly resolved neutron spectrum.

The Schr\"odinger problem of a neutron submitted to the 
local gravitational potential, $V=mgx$ has well known solutions
(here $m$ is the neutron's gravitational mass and 
$x$ the vertical height). The eigen-solutions of 
$\hat{H} \Psi_n = E_n  \Psi_n$, 
can be expressed in terms of Airy functions, $\Phi$, \cite{Landau}:
\beq
\Psi_n(z) = A_n ~\Phi(z)~~,
\label{airy}
\eeq
with eigenvalues determined by the roots of the Airy functions:
\beq
\alpha_n = - \left({2 \over m g^2 \hbar^2}\right)^{1/3} E_n ~~.
\label{airyzero}
\eeq
The variable $z$ has a simple relation to the physical height $x$ which for the
Nesvizhevsky et al. experiment is of order $\mu m$:
\beq
z = \left({2 m^2 g \over \hbar^2}\right)^{1/3}(x - {E_n \over mg})~~. 
\label{z}
\eeq
The normalization of the wavefunction $A_n$ can be explicitly determined
in terms of the integral
\beq
A_n^{-2} =  \left({\hbar^2 \over 2 m^2 g}\right)^{1/3}
\int_{-\alpha_n}^{\infty} \Phi(z)^2 ~ dz~~.
\label{norm}
\eeq 
First of all, directly from Eq. (\ref{airyzero}), one can extract a precision
for the measurement of the local gravity of 
$\Delta g = 6 \times 10^{-9}$ ms$^{-1}$,
under the assumption that $\Delta E = 10^{-18}$ eV.
We will show that other effects superpose this uncertainty.

Let us then establish to which extent the Weak Equivalence Principle 
can be tested 
once the energy resolution of the Nesvizhevsky et al. experiment is improved.
For that one has to realize that in the neutron's Hamiltonian two masses are 
present. The neutron's inertial mass, $m_i$, in the kinetic term and the 
gravitational mass 
in the gravitational potential. Distinguishing these two masses implies the 
relationship (\ref{airyzero}) is slightly altered as $m$ must be replaced 
by $m^2/m_i$. It then follows that the corresponding uncertainty in the energy 
due to the difference between inertial and gravitational masses is given by 

\beq
{m_i - m \over m} = {3 \over 2} m g^2 \hbar^2 \alpha_n^3 
{\Delta E \over E_n^4}~~.
\label{deltaenergy}
\eeq

\noindent
Hence, for the ground state, $E_0 = 1.44~peV$ \cite{Nesvizhevsky1} and 
$\alpha_0=2.338$, and if $\Delta E=10^{-18}~eV$ 
is attained, one can
establish a somewhat modest bound:
\beq
{m_i - m \over m} < 1.9 \times 10^{-6}~~.
\label{gravinertial}
\eeq

The situation is somewhat different in what concerns a new force of Nature.
The exciting prospect of a new fundamental interaction,
beyond the already known four interactions, sparked in 1986 
from the claim that the original \"Eotv\"os experiment, 
designed to verify the equality of gravitational and inertial masses, 
revealed evidence of a new force with sub-gravitational strength (see Ref. 
\cite{Fishbach1} and references therein for an exhaustive discussion). 
Despite the fact that \"Eotv\"os data turned out not be conclusive, 
the claim has stimulated a great deal of theoretical discussion 
(see e.g. \cite{Nieto} for a complete set of references) as well as the 
repetition of old experiments using new technology. Actually, 
the most stringent limit on the equality of inertial and gravitational 
masses (of $Cu$ and $Pb$) has arisen in this context and  
with an accuracy of $5 \times 10^{-13}$ \cite{Adelberger}.

The simplest way a new interaction or a {\it fifth force} could arise 
would be through the exchange of a light boson coupled to matter 
with sub-gravitational strength. 
This could originate from various physical models at the Planck scale 
such as the extended supergravity theories after dimensional reduction 
\cite{Nieto,Scherk}, the compactification of $5$-dimensional
generalized Kaluza-Klein theories that include gauge interactions at
higher dimensions \cite{Bars}  and also from string/M-theory. 
A common feature of these schemes is the appearance of a new Yukawa type 
modification in the interaction energy, $V(r)$, between two 
point masses $m_1$ and $m_2$:
\begin{equation}
V(r) = -  {G_{\infty}~m_{1}~m_{2} \over r}~
[1 + \alpha_5~\exp{(-r/\lambda_5})]~~,
\label{eq:1.1}
\end{equation}
where $r = \vert \vec r_{2} - \vec r_{1} \vert$ is the distance between the
masses, $G_{\infty}$ is the gravitational coupling for $r \rightarrow \infty$, 
$\alpha_5$ and $\lambda_5$ are the strength and the range of the new 
interaction.
Of course, $G_{\infty}$ is related with the Newtonian gravitational 
constant. Indeed, the force associated with Eq. (\ref{eq:1.1}) is given by:
\begin{equation}
\vec F(r) = -  \nabla V(r) = - {G(r)~m_{1}~m_{2} \over r^2}~\hat {\bf r}~~,
\label{eq:1.2}
\end{equation}
where
\begin{equation}
G(r) = G_{\infty}[1 + \alpha_5~(1 + r/\lambda_5)~\exp{(-r/\lambda_5)}]~~.
\label{eq:1.3}
\end{equation}

A particularly interesting implication of the mentioned approaches to 
the Planck scale physics is that 
the coupling  $\alpha_5$ is not an universal constant, 
but instead a parameter dependent on the 
chemical composition of the test masses as first pointed out in 
Ref. \cite{Lee}. 
This dependence comes about if one assumes that the  new
bosonic field couples to the baryon number $B = Z + N$ (the sum
of protons and neutrons) and would imply a clear violation of the 
Weak Equivalence Principle.

Several experiments have been performed in order to establish the 
parameters of a new interaction based on the idea of a composition-dependence 
differential free fall of bodies (see Refs. \cite{Fishbach1,Nieto} 
for discussions on those leading to the most stringent limits). 
The current data is entirely 
compatible with predictions of Newtonian gravity from 
both, composition-independent or composition-dependent, experiments.
The  bounds on parameters $\alpha_5 $ and $\lambda_5$ can be summarized as 
follows:

\begin{itemize}

\item
Laboratory experiments devised to measure deviations from 
the inverse-square law are sensitive to the range 
$10^{-2}~m < \lambda_5 \lsim 1~m$ and constrain $\alpha_5$ to be smaller than 
$10^{-4}$;

\item
Nucleosynthesis bounds imply that $\alpha_5 \lsim 4 \times 10^{-1}$ for 
$\lambda_5 \lsim 1~m$ \cite{Bertolami2};

\item
Gravimetric experiments sensitive in the range of
$10~m \lsim \lambda_5 \lsim 10^{3}~m$ suggest $\alpha_5 < 10^{-3}$;

\item
Satellite tests probing ranges about $10^{5}~m \lsim \lambda_5 \lsim 10^{7}~m$ 
indicate that $\alpha_5 < 10^{-5}$;

\item 
Radiometric data of the Pioneer 10/11, Galileo and Ulysses 
spacecrafts suggest the presence of a new force with parameters 
$\alpha_5 = - 10^{-3}$ and $\lambda_5 \simeq 4 \times 10^{13}~m$ 
\cite{Anderson}.

\end{itemize}

It is striking that, for $\lambda_5 < 10^{-3}~m$ and  
$\lambda_5 > 10^{13}~m$, $\alpha_5$ is essentially unconstrained. 
The former range arises in higher dimensional superstring motivated 
cosmological solutions, where matter fields that are related to open 
string modes, lie on a lower dimensional brane, while gravity propagates 
in the bulk \cite{Polchinski}. In these scenarios, the $d$ extra dimensions 
are not restricted to be small \cite{Antoniadis} and 
the fundamental $D$-dimensional scale, $M_D$, with $D=4+d$, can be 
considerably smaller than the 4-dimensional Planck scale.
Assuming that these quantum gravity effects are just beyond current 
detection capability in accelerators, then $M_D \gsim few~\tev$ 
and hence modifications 
to Newtonian gravity will occur in the short range region, 
$\lambda_5 < 10^{-3}~m$. This range also emerges if one assumes 
that scalar \cite{Beane} or vector/tensor \cite{Bertolami3} 
excitations are associated with the observed vacuum energy density.

As a matter of fact, the millimeter range has been recently 
available for experimental verification and as a consequence the effects 
associated to $d$ extra dimensions, that imply that $\alpha_5 = d+1~(2d)$ 
for a circle 
(torus) topology \cite{Floratos}, have been ruled out down to
$\lambda_5 \lsim 0.2~mm$ yielding, for two extra dimensions, the bound 
on the typical energy scale, $M_6 \ge 3.5~\tev$ \cite{Hoyle}. 
  
It should be mentioned that cosmological considerations tend to 
yield higher bounds for $M_D$. 
For instance, requiring that nucleosynthesis yields are not 
affected by brane related effects, imply that 
$M_5 \simeq 10~\tev$ \cite{Cline} and hence 
$\lambda_5 < 10^{-4}~m$. On the other hand, 
matching inflationary observable 
quantities in the context of supergravity models on the 
brane, yield much higher bounds for $M_5$, 
typically, $M_5 \gsim 10^{13}~\gev$ for supergravity mass term 
chaotic inflation 
\cite{Bento1} and $M_5 \lsim 10^{16}~\gev$ for supergravity inflationary 
models where the potential has the form $V = V_0 [1 + c_n (\phi / M_P)^n]$, 
the first term being dominant \cite{Bento2}.

Let us examine to which extent the Grenoble experiment 
can allow imposing limits 
on the parameters of a putative new interaction of Nature once the energy 
resolution 
of the experiment is increased to $\Delta E=10^{-18}~eV$.

In order to study possible limits on a {\it fifth force} which is sensitive 
to baryon number, we assume that its effects can be treated as a perturbation
of the local gravitational potential.
In this way, first order perturbation theory
provides a correction to the energy of the n$^{th}$-level defined by:
\beq 
\Delta E_n^{(1)} = \int_0^{\infty} \Psi_n^* ~ V(x) ~ \Psi_n~~ dx,
\label{pertenergy}
\eeq
where ${\Psi_n}$ is the unperturbed eigenstate and
the perturbation potential is given by 
$V(x) = \alpha_5 f ~ mg x  \exp{(-x/\lambda_5)}$, where $f$ is the 
fraction of the source (Earth) that contributes to the force acting on the 
neutrons.
Since the neutrons are constrained vertically in the length scale
$ x \approx \mu m$, the bounds on $\alpha_5 f$ become interesting
for say $\lambda \gsim 10 \mu m$. In this range, the exponential in 
$V(x)$ can be expanded around $x=0$. Retaining 
the first two terms in the expansion, we derive  an upper bound for 
$\alpha_5 f$ in terms of the energy resolution $\Delta E$ of the 
experiment: 
\beq
\alpha_5 f \left[3.24 + 5.05 \left({\lambda_5 \over 1~\mu m}\right)
^{-1}\right] 
\lsim {4.91 \times 10^{-2}} \left({\Delta E \over 10^{-18}~eV}\right).
\label{alfalam}
\eeq  
Here, we have used the numerical values for the
following integrals 
\beq
I_0 =  \int_{-\alpha_0}^{\infty} \Phi(z)^2 ~ dz = 1.24 \times 10^{-4}~~,
\label{integral0}
\eeq
\beq
I_1 =  \int_{-\alpha_0}^{\infty} \Phi(z)~ z ~\Phi(z)~ dz = 3.24 
\times 10^{-3}~~,
\label{integral1}
\eeq
\beq
I_2 =  \int_{-\alpha_0}^{\infty} \Phi(z) ~z^2 ~\Phi(z) ~ dz= 8.64 
\times 10^{-4}~~.
\label{integral2}
\eeq

Directly from Eq. (\ref{alfalam}), we conclude that, within the range 
where no constraints exist $10 \mu m \lsim \lambda \leq 1 cm$, 
the upper limit on $\alpha_5 f$
would be of the order of  $ 1.3 \times 10^{-2}$ as long
as the energy resolution of $\Delta E = 10^{-18}~eV$ is achieved. 
Since the tested region for the range is quite small, $f << 1$ implying 
that there is no limit for $\alpha_5$ as the effect of a possible 
new force is far too small to affect neutrons in the experiment. 
If on the other hand the new force 
were long range so that $f=1$, still no bound on $\alpha_5$ would arise 
given the relationship between Newton's constant, $G_N$ and $G_{\infty}$, 
namely $G_N = G_{\infty} (1 + \alpha_5)$. We mention that the same 
conclusions can be drawn for the so-called massive version of the 
Brans-Dicke theory \cite{Acharya}, where 
$\lambda_5 = m_{\phi}^{-1}$ and $\alpha_5 = 1/2 \omega + 3$, $m_{\phi}$ 
being the scalar field mass and $\omega$ the Brans-Dicke coupling parameter.

We turn now to the issue of the gravitational screening. 
A possible 
absorption of the gravitational force between two bodies, when a medium 
is screened by another, has been the subject of a series of tests. 
This effect is reminiscent 
of the magnetic permeability of materials, and a screening or extinction 
coefficient, $h$, was proposed by Quirino Majorana \cite{Majorana} 
in 1920 in order 
to measure the ability of an object of dimension $L$ 
with density $\rho(r)$ to shield the 
gravitational force between masses $m_1$ and $m_2$:
\beq 
F^{\prime}  = {G m_1 m_2 \over r^2} 
\exp\left[- h \int_0^L \rho(r)~dr \right]~~.
\label{eq:2.10}
\eeq

\noindent
Naturally, $h$ must be quite small. Several attempts to measure this 
constant from general principles have been made. For instance, 
Weber \cite{Weber} has argued that 
quasi-static shielding could be predicted from a general relativistic 
analysis of tidal phenomena, and stated that the effect should be 
extremely small. More recently, it has been shown that a lunar laser 
ranging experiment can set the impressive limit, 
$h \leq 1.0 \times 10^{-21}~cm^2 g^{-1}$ \cite{Eckhardt}. 
The most stringent laboratory limit on the gravitational shielding 
constant has its origin on a recent measurement of Newton's 
constant carried out 
at the Physik-Institut, Universit\"at Z\"urich, which yields  
$h \leq 4.3 \times 10^{-14}~cm^2 g^{-1}$ \cite{Unnikrishnan}. 

In order to use an improved version of the Grenoble experiment for 
obtaining a limit on the gravitational screening one considers
the effect of the Moon on the local gravity acceleration.
This effect is about $g_M \simeq 3 \times 10^{-6}~g$. Of course, 
an experiment of this nature implies accurate estimates 
of various local gravitational  perturbations 
related to human activity and geophysical nature. 
Comparing the contribution of the Moon when Moon, Earth and the experiment are 
in alignment and the former is ``screened'' by Earth, then 
$g_{M2} = 3.3 \times  10^{-6}~g$, from which follows that an effect on the 
neutron spectra will be felt only if:
\beq
h > 1.7 \times 10^{-10} cm^2~g^{-1}~~.
\label{screening}
\eeq
Here we have assumed that Earth's density is constant over its diameter: 
$\rho_{\oplus} = 5.51~g~cm^{-3}$. The result in Eq. (\ref{screening})
is much less stringent than the abovementioned bounds. Similar analysis 
using the Sun leads to a limit that is about $4$ times greater.

Naturally, any actual experiment to measure gravitational origin effects on 
ultracold neutrons requires besides the six orders of magnitude 
improvement in the 
energy resolution of the neutron spectrum, a series of precautions 
in order to eliminate local gravitational perturbations and asymmetries.

\begin{acknowledgments}

\noindent
The authors would like to thank Clovis de Matos for  
discussions and Prof. Edward Witten for valuable comments.

\end{acknowledgments}


\begin{thebibliography}{99}



\bibitem{Nesvizhevsky} V.V. Nesvizhevsky et al., \NAT {\bf 415} (2002) 297.


\bibitem{Luschikov} V.I. Luschikov and A.I. Frank, {\it JEPT} {\bf 28} 
(1978) 559.


\bibitem{Nesvizhevsky1} V.V. Nesvizhevsky et al.,  {\it Nucl. Instruments 
and Methods in Physics Research} {\bf A440} (2000) 754.


\bibitem{Collela} R. Collela, A.W. Overhauser and S.A. Werner, \PRL {\bf 34} 
(1975) 1472.


\bibitem{Bertolami1} O. Bertolami, \MPL {\bf A6} (1986) 383.


\bibitem{Landau} L. Landau and E. Lifchitz, ``M\'ecanique Quantique'' 
(Mir, Moscou 1967).


\bibitem{Fishbach1} E. Fishbach and C. Talmadge, ``Ten Years of the Fifth  
Force'', hep-ph/9606249.


\bibitem{Nieto} M.M. Nieto and T. Goldman, \PRTS {\bf 205} (1991) 221.
 

\bibitem{Adelberger} G.L. Smith et al., \PR {\bf D61} (2000) 022001.


\bibitem{Scherk} J. Scherk, \PL {\bf B88} (1979) 265. 


\bibitem{Bars} I. Bars and M. Visser, \PRL {\bf 57} (1986) 25. 


\bibitem{Lee} T.D. Lee and C.N. Yang, \PR {\bf 98} (1955) 1501.


\bibitem{Bertolami2} O. Bertolami and F.M. Nunes, \PL {\bf B452} (1999) 108. 


\bibitem{Anderson} J.D. Anderson et al., \PR {\bf D65} (2002) 082004.


\bibitem{Polchinski} J. Polchinski, \PRL {\bf 75} (1995) 4724;

P. Horava and E. Witten, \NP {\bf B460} (1996) 506;

A. Lukas, B.A. Ovrut and D. Waldram, \PR {\bf D60} (1999) 086001.


\bibitem{Antoniadis} I. Antoniadis, \PL {\bf B246} (1990) 317;

N. Arkani-Hamed, S. Dimopoulos and G. Dvali, \PL {\bf B429} (1998) 263.


\bibitem{Beane} S.R. Beane, \GRG {\bf 29} (1997) 945.


\bibitem{Bertolami3} O. Bertolami, \CQG {\bf 14} (1997) 2785.

 
\bibitem{Floratos} E.G. Floratos and G.K. Leontaris, \PL {\bf B465} (1999) 
95.
 

\bibitem{Hoyle} C.D. Hoyle et al.,  \PRL {\bf 86} (2001) 1418.


\bibitem{Cline} J.M. Cline, C. Grojean and G. Servant, \PRL {\bf 83} (1999) 
4245.


\bibitem{Bento1} M.C. Bento and O. Bertolami, \PR {\bf D65} (2002) 063513.


\bibitem{Bento2} M.C. Bento, O. Bertolami and A.A. Sen, 
``Supergravity Inflation on the Brane'', gr-qc/0204046.


\bibitem{Acharya} R. Acharya and P.A. Hogan, \LNC {\bf 6} (1973) 668.


\bibitem{Majorana} Q. Majorana, \PMAG {\bf 39} (1920) 488.



\bibitem{Weber} J. Weber, \PR {\bf 146} (1966) 935.



\bibitem{Eckhardt} D.H. Eckhardt, \PR {\bf D42} (1990) 2144.



\bibitem{Unnikrishnan} C.S. Unnikrishnan and G.T. Gillies, 
\PR {\bf D61} (2000) 101101(R).




\end{thebibliography}
\end{document}